\documentclass[aps,prl,twocolumn,preprintnumbers]{revtex4}
\usepackage{epsfig}

\newcommand{\MS}{\overline{\rm MS}}

\newcommand{\be}{\begin{equation}}
\newcommand{\ee}{\end{equation}}
\newcommand{\bea}{\begin{eqnarray}}
\newcommand{\eea}{\end{eqnarray}}
\def\lQ{\Lambda_{\rm QCD}}
\def\als{\alpha_{\rm s}}
\newcommand{\nn}{\nonumber}

\newcommand{\Eqn}[1]{Eq.~(\ref{#1})}

\newcommand{\cO}{{\cal O}}

\begin{document}

\preprint{UB-ECM-PF-06-04; IPPP/06/05}

\title{\bf Renormalization Group Improved Sum Rule Analysis 
for the Bottom Quark Mass} 
\author{\bf Antonio~Pineda$^1$ and  Adrian~Signer$^2$}
\affiliation{
$^1\!\!$ Dept. d'Estructura i Constituents de la Mat\`eria,
U. Barcelona, Diagonal 647, E-08028 Barcelona, Catalonia, Spain\\
$^2\!\!$ Institute for Particle Physics Phenomenology,
Durham, DH1 3LE, England }

\date{23 January 2006}

\begin{abstract} 
\noindent
We study the effect of resumming large logarithms in the determination
of the bottom quark mass through a non-relativistic sum rule analysis.
Our result is complete at next-to-leading-logarithmic accuracy and
includes some known contributions at next-to-next-to-leading
logarithmic accuracy. Compared to finite order computations, the
reliability of the theoretical evaluation is greatly improved,
resulting in a substantially reduced scale dependence and a faster
convergent perturbative series. This allows us to significantly
improve over previous determinations of the $\MS$ bottom quark mass,
$\overline{m}_b$, from non-relativistic sum rules. Our final figure
reads $\overline{m}_b(\overline{m}_b)=4.19\pm 0.06$~GeV.
\end{abstract}
\maketitle

Processes involving a $b\bar{b}$ quark pair close to threshold are
very sensitive to the bottom quark mass, $m_b$, and offer a unique
opportunity to accurately determine its value. One of the cleanest
observables where this dependence on $m_b$ shows up is the
non-relativistic sum rule~\cite{Novikov:1976tn}
\begin{equation}
\label{momdef}
M_n \equiv \frac{12\pi^2 e_b^2}{n!} 
\left(\frac{d}{dq^2}\right)^n \Pi(q^2)\big|_{q^2=0} = 
\int_{0}^\infty \frac{ds}{s^{n+1}} R_{b\bar{b}}(s) , 
\end{equation}
where $R_{b\bar{b}}(s)\equiv \sigma(e^+e^-\to b\bar{b})/
\sigma(e^+e^-\to \mu^+\mu^-)$, $\Pi(q^2)$ is the vacuum polarization
and $e_b=-1/3$ the electric charge of the bottom quark.  The typical
scale is $p\sim 2 m_b/\sqrt{n}$ and, provided $n$ is not chosen to be
too large, the left-hand side of Eq. (\ref{momdef}) can be reliably
computed using a weak coupling analysis (the right-hand side can be
determined from experiment).  To describe such processes
theoretically, a standard fixed-order calculation in the strong
coupling $\als$ is insufficient due to the presence of terms
$(\als/v)^n\sim 1$ at each order in perturbation theory, where $v \sim
1/\sqrt{n}\ll 1$ is the velocity of the heavy quarks. Such terms
appear because there are several scales involved in the problem. There
is the hard scale $\mu_h\sim m_b$, the soft scale $\mu_s\sim m_b v$ of
the order of the typical momentum and, finally, the ultrasoft scale
$\mu_{us}\sim m_b v^2$ of the order of the typical kinetic energy of
the heavy quarks. Using effective field theories (for a review see
\cite{Brambilla:2004jw}), the perturbative expansion can be
systematically reorganized into an expansion in the two small
parameters of the problem, $\als$ and $v$, and the $b\bar{b}$ cross
section can be written as
\begin{eqnarray}
\label{RbbLO}
&&R_{b\bar{b}}(s) = v \sum_{n} \left(\frac{\als}{v}\right)^n
\\
\nn
&&
\times
\left\{ 1 ({\rm LO}); \als, v ({\rm NLO}); \als^2, v \als,
 v^2 ({\rm NNLO}) \ldots \right\}.
\end{eqnarray}
The coefficients of this series can be computed most efficiently using
the threshold expansion \cite{Beneke:1997zp}.

At present, non-relativistic sum rules have been computed at
next-to-next-to-leading order (NNLO)~\cite{nnlo,Beneke:1999fe}.  This
allowed for a precise determination of the bottom quark mass using a
well understood perturbative approach. 
Unfortunately, in the on-shell scheme, the NNLO
corrections turned out to be much larger than anticipated and,
moreover, very strongly scale dependent.
The use of threshold masses~\cite{massdef,Beneke:1998rk,Pineda:2001zq}, 
which account for the
cancellation of the pole mass renormalon in the observable, do not really solve these problems, specially for the strong scale dependence.
Overall, this produced a very slowly
convergent series, being the dominant source of error in the
determination of $m_b$.  Non-perturbative corrections are known in the
limit $m_b/n \gg \lQ$ \cite{Voloshin:1995sf}.  Even though this limit
does not hold for large enough $n$, we can take it as an order of
magnitude estimate. Numerically, these corrections are very small and
can be neglected in comparison with other sources of errors.

The situation is very similar in the case of $t\bar{t}$ pair
production near threshold. In this case the use of threshold masses
results in a well behaved perturbative series for the position of the
peak of the $t\bar{t}$ cross section and, therefore, may enable a
precise determination of the top quark mass, once experimental data is
available. However, the large theoretical uncertainty in the
normalization of the cross section remained (even if the series is
more convergent than in the bottom case). It has been claimed
\cite{Hoang:2000ib} that the source of this uncertainty is due to
potentially large $\log v$ terms, which arise due to the presence of
several scales and take the form $\log \mu_h/\mu_s$ and $\log
\mu_s/\mu_{us}$. Although it has been shown later
\cite{Pineda:2001et,Pineda:2001ra} that some expressions used in that
paper were incorrect and that the final error claimed was somewhat
over optimistic~\cite{Hoang:2003ns}, the message that the resummation
of logarithms improves the scale dependence has gone through.

Given the importance of the $\log v$ terms for the $t\bar{t}$ cross
section, it is natural to ask whether their inclusion also improves
the situation in the $b\bar{b}$ case.  In our case we have to replace
the expansion of \Eqn{RbbLO} by \bea
\label{RbbLL}
&&
R_{b\bar{b}}(s) = v \sum_{n} \left(\frac{\als}{v}\right)^n
\sum_{m} \left( \als \log v\right)^m 
\\
\nn
&&
\times
\left\{ 1 ({\rm LL}); \als, v ({\rm NLL}); \als^2, v \als,
 v^2 ({\rm NNLL}) \ldots \right\}
\,.
\eea
As we will see, these logarithms are extremely important numerically and
substantially improve the reliability of the theoretical evaluation of
the moments.

The $n$-th moment, $M_n$, as defined in \Eqn{momdef} is computed in
the usual way. First we match QCD to non-relativistic QCD (NRQCD) at
the hard scale which we set to $\mu_h=m$. This theory is then matched
to potential NRQCD (pNRQCD) \cite{Pineda:1997bj}. Solving the
corresponding non-relativistic Schr\"odinger equation perturbatively
we obtain ${\rm Im}\, G(0,0,E)$, the imaginary part of the Green
function at the origin. $M_n$ can then be written as
\bea
\label{Mnexpr}
M_n &=& 48 \pi e_b^2 N_c 
\int_{-\infty}^\infty \frac{dE}{(E+2m_b)^{2n+3}}
\\
\nn
&&
\times\left(c_1^2 - c_1 c_2 \frac{E}{3m_b}\right) 
{\rm Im}\, G(0,0,E) , 
\eea
where $E=\sqrt{s}-2 m_b$, $N_c=3$ and $c_1$ and $c_2$ are the matching
coefficients of the currents, normalized to 1 at leading order. In a
strict non-relativistic expansion one also expands
\begin{equation}
\label{emExp}
\frac{1}{(E+2m_b)^{2n+3}} = 
 \frac{e^{-n \frac{E}{m_b}}}{(2m_b)^{2n+3}} 
 \left( 1-\frac{3E}{2m_b}+ \frac{n E^2}{4 m_b^2} \ldots \right)
\end{equation}
treating $n E \sim m_b$. We also remark that the logarithms involving
$\mu_s$ always appear in the combination $\log(-4m_b E/\mu^2_s)$. This
confirms that the natural scales are given by $E \sim m_b/n$ and
$\mu_s\sim p\sim 2 m_b/\sqrt{n}$. To ensure the applicability of
perturbation theory, we cannot choose $n$ too large and will restrict
ourselves to $n\le 14$.

The matching coefficients of pNRQCD depend on the scales $\mu_h=m_b$,
$\mu_s$ and $\mu_{us}$. In solving the renormalization group equations
we have set $\mu_{us} = \mu_s^2/m$.  The expressions we use are
complete at NLL and NNLO. At NNLL they are also complete (in
particular we include the insertions of the renormalization group
improved potentials to $G(0,0,E)$ up to the desired order in the $\MS$
scheme) except for $c_1$.  For $c_1$ we are using the known
NLL~\cite{Pineda:2001et} expression as well as some partial NNLL
contributions, which include the spin-dependent
corrections~\cite{Penin:2004ay}, the NNLL ultrasoft corrections to the
Green function, the corrections due to the two-loop beta running, and
some contributions coming from the introduction of partial
higher-order terms in the renormalization group improved potentials
that appear in the anomalous dimension of $c_1$. For details we refer
to Refs.~\cite{Beneke:1999fe, longP}.  In particular we stress that
not all the ultrasoft related logarithms are included in our
analysis. With this caveat in mind, we still refer to our full result
as NNLL.

We also include QED corrections in our result. Counting $\alpha \sim
\als^2$, these corrections enter already at NLO, due to a single
exchange of a potential photon, but they have only a minor numerical
impact. They increase the extracted bottom quark mass by less than
10~MeV. 

The threshold masses we consider in this analysis are the potential
subtracted (PS) mass $m_{b,{\rm PS}}(\mu_f)$~\cite{Beneke:1998rk} and
the renormalon subtracted (RS) mass $m_{b,{\rm
RS}}(\mu_f)$~\cite{Pineda:2001zq}. The subtraction scale $\mu_f$ that
is needed for the definition of the PS/RS mass is set to $\mu_f =
2$~GeV, to ensure it does not exceed the characteristic scale
$\mu_s$. Once the PS/RS mass is determined, we convert it to
$\overline{m}_b$, the $\overline{\rm MS}$ mass at the renormalization
scale $\overline{m}_b$. We use the three-loop
conversion~\cite{Chetyrkin:1999} of the pole mass to $\overline{m}_b$
and for the PS and RS mass a `large $\beta_0$' \cite{Ball:1995ni} and
renormalon-based \cite{Pineda:2001zq} approximation respectively for
the four-loop term.

The moments are evaluated by performing the energy integration in the
complex energy plane using a strictly expanded form as indicated in
\Eqn{emExp}. The difference between this evaluation and using 
\Eqn{Mnexpr} is NNNLO and,
therefore, beyond the accuracy we are aiming at. However, for small
values of $n$, this difference is sizable. In fact, for $n=6$ the
resulting values for $m_{b,{\rm PS/RS}}$ may differ by up to 45/60~MeV 
depending on how we expand the prefactor \Eqn{emExp},
for $n=8$ the difference is up to 15/25~MeV, whereas for $n\ge10$ the
values for $m_b$ agree within 10/15~MeV.

The experimental moments are determined as described in
Ref.~\cite{Beneke:1999fe}. The moment is split into the contribution
due to the six $\Upsilon$ resonances and the open $b\bar{b}$
continuum. The main uncertainty comes from the rather poor knowledge
of the latter, which we parametrize as $R_{b\bar{b}}^{cont} = 0.4\pm
0.2$~\cite{Akerib:1991eq}. Since the continuum contribution is
suppressed for larger values of $n$, resulting in a smaller
experimental error, we refrain from using $n < 6$.

\begin{figure}
   \epsfxsize=9cm
   \centerline{\epsffile{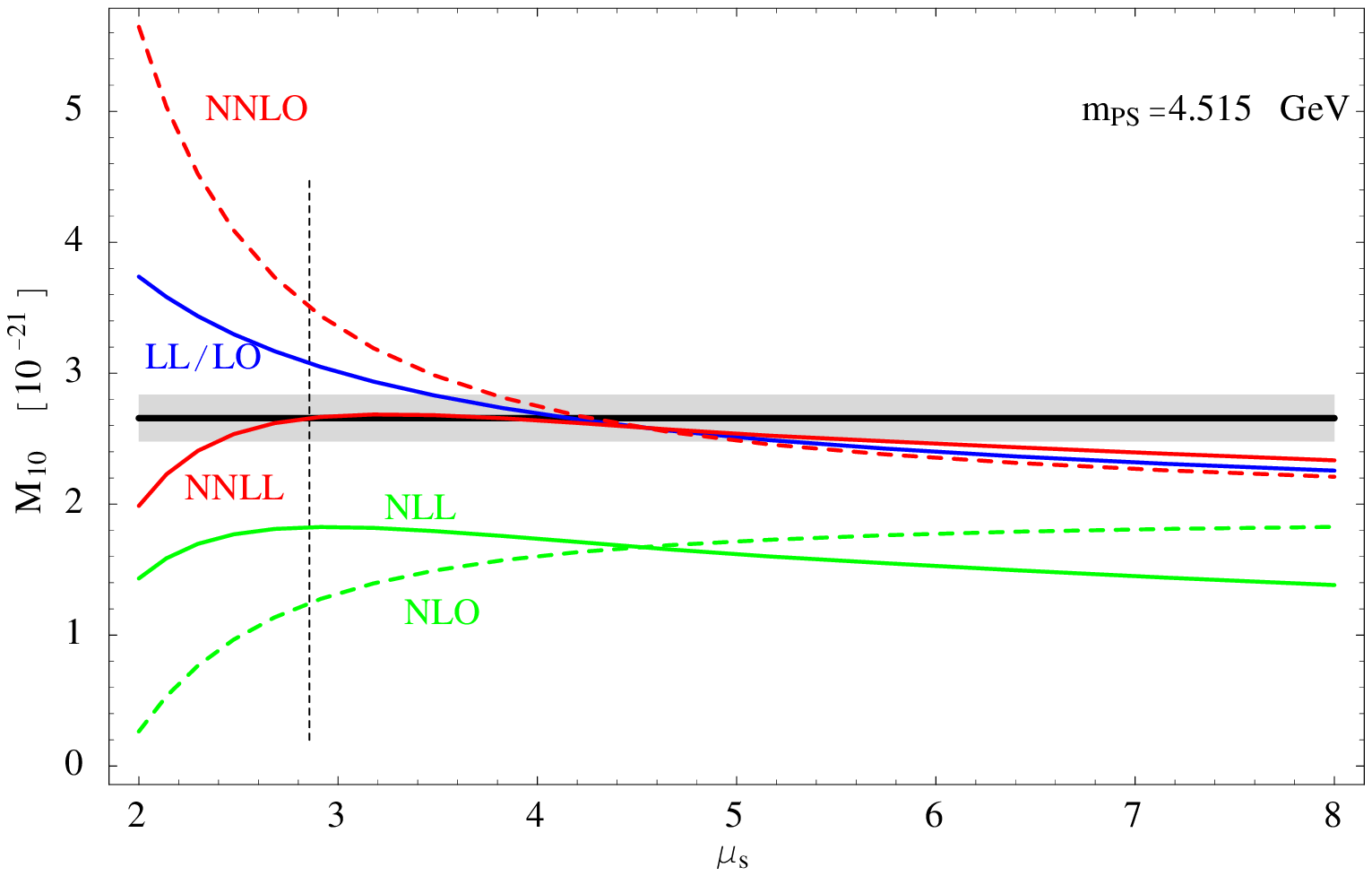} }
   \centerline{\epsffile{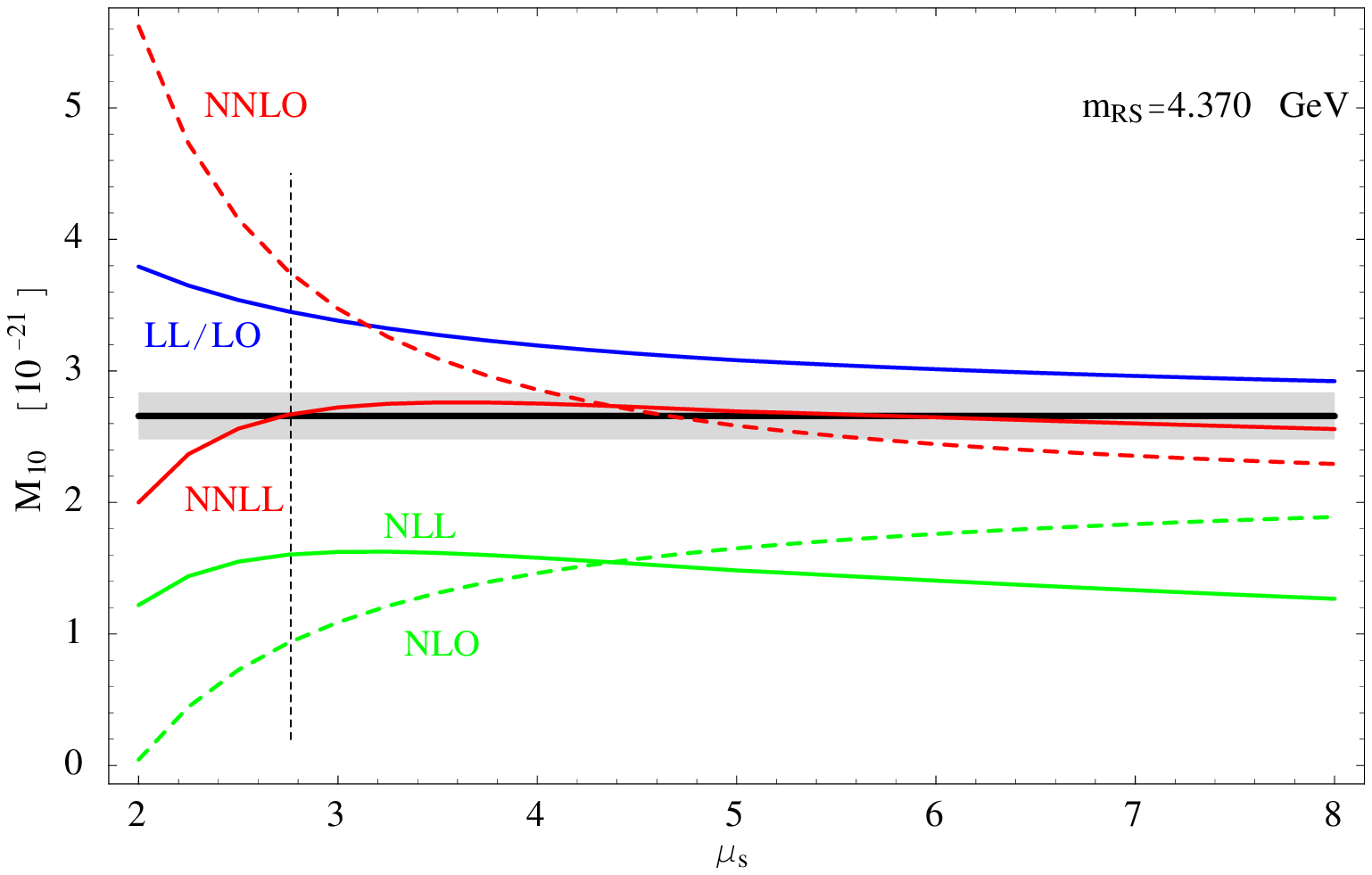} }
   \vspace*{0.2cm} \caption{\it The moment $M_{10}$ as a function of
 $\mu_s$ at LO/LL, NLO, NLL, NNLO and NNLL for $m_{b {\rm PS}}$(2~GeV)
 = 4.515~GeV in the PS scheme (upper figure), and for $m_{b {\rm
 RS}}$(2~GeV) = 4.370~GeV in the RS scheme (lower figure). The
 experimental moment with its error is also shown (grey band).
 \label{fig:Mom10}}
\end{figure}

The main theoretical uncertainty in previous determinations of the
bottom quark mass was due to the huge scale dependence of the NNLO
result, which made it rather difficult to find a reliable procedure
for estimating the theoretical error. It is the main result of this
work to show that the situation improves considerably if a
renormalization group improved analysis is performed. To illustrate
this, in Figure~\ref{fig:Mom10} we show the dependence of the
theoretical value for $M_{10}$ (evaluated at LO/LL, NLO, NLL, NNLO and
NNLL respectively) on $\mu_s$. For the purpose of illustration we also
plot the experimental value of the moment including its error. We set
the strong coupling to $\als(M_Z)=0.118$ and use three-loop evolution
to determine it at lower scales. For the plot shown in
Figure~\ref{fig:Mom10} we set $m_{b, {\rm PS}}$(2~GeV) = 4.515~GeV and
$m_{b, {\rm RS}}$(2~GeV) = 4.370~GeV and vary the soft scale around its
characteristic value $\mu_s\sim 2 m_b/\sqrt{10}$ (indicated by a
dashed vertical line). Note that in this region the size of the NNLL
corrections (even if large) is considerably smaller than the
corresponding fixed-order NNLO ones.  Moreover, contrary to earlier
analyses, the NNLL result is now more stable with respect scale
variations than the NLL one (for the range of scales for which the
computation is trustworthy). This is of course what one would expect
and indicates that the inclusion of the logarithms substantially
improves the reliability of the theoretical prediction. Only for
scales $\mu_s< 2$~GeV the situation gets out of control, but for these
scales the ultrasoft scale is below 1 GeV and we can not really rely
on our computation. Multiple insertions of corrections to the Coulomb
potential also seem to be important in this region
\cite{Penin:2005eu}.

The situation is similar for other values of $n$. As a general
feature, for increasing $n$, the scale dependence increases slightly. 
This is not surprising since larger
$n$ induce smaller scales and at some point the applicability of
perturbation theory is questionable. On the other hand, as mentioned above,
smaller values of $n$ have the disadvantage that the non-relativistic
approximation becomes less reliable.

These findings show that it is possible to improve the accuracy of
previous determinations of the bottom quark mass from non-relativistic
sum rules if the renormalization group improvement is applied. In
order to determine the $\MS$ mass we first determine the PS/RS mass
with its error, proceeding as follows: we consider $M_n$ for
$n\in\{6,8,10,12,14\}$ and obtain our central value by equating the
theoretical and experimental value of the moment at the standard scale
$\mu_s=2 m_b/\sqrt{n}$. For the error in the determination of the
threshold masses we consider three sources: the experimental error,
the error due to the uncertainty in the strong coupling and finally
the theoretical error.

The experimental error, $\Delta_{\rm exp}$, is simply determined by
extracting the value for $m_b$ for the two extreme values of the
experimental moment. The error due to the uncertainty in the strong
coupling, $\Delta_\alpha$, is obtained by studying the effect on the
extracted bottom quark mass if we vary $0.115 < \als(M_Z) < 0.121$.
Following common practise one would estimate the theoretical error,
$\Delta_{\rm th}$, by variation of the scale by a factor of two. As is
obvious from Figure~\ref{fig:Mom10}, for small scales the theoretical
result cannot be trusted. Therefore, in previous analyses, the scale
variation was limited to scale choices above a certain cutoff,
typically set to a value around 2~GeV. In the current analysis we
refrain from using such an estimate. There are several reasons. First,
such an error estimate depends crucially on the somewhat arbitrary
lower cutoff of the scale variation. Second, it does not take into
account the fact that the higher-order corrections are sizable. Given
that the scale dependence is very modest (for reasonably large scales)
compared to the size of the NNLL corrections, we think that such an
error analysis would considerably underestimate the theoretical error
in the present case. Finally, the scale variation as depicted in
Figure~\ref{fig:Mom10} does not take into account the independent
variation of the ultrasoft scale, since in our analysis the latter is
determined by the soft scale. It would be preferable to be able to
vary all scales independently to obtain a better estimate of the
uncertainty, in particular since some ultrasoft logarithms are missing
in our result. We have verified that a naive variation of $\mu_{us}$
results in a rather large uncertainty which, however, is consistent
with the final error estimate we propose. Therefore, we prefer to
determine the theoretical error by taking half the size of the
highest-order correction that is included in our result. More
precisely, we determine two values for $m_b$ by equating the
experimental and theoretical value (at the scale for which it reaches
its maximum) of the moment at NNLL and NLL respectively. The error is
determined as half the difference between these two values. 
This procedure assumes a perturbative series where successive terms
become less and less important. For this to hold we have to use a
threshold mass, since for the pole mass the NNLL corrections are much
larger than the NLL ones. In this respect moments with low values of
$n$ and/or threshold mass definitions with values close to the $\MS$
mass are better behaved. On the other hand, the actual size of the
correction, and therefore the assigned error, increases for such mass
definitions.

\begin{table}[h]
\begin{center}
\begin{tabular}{|c|c|c|c|c|c|c|}
\hline
$n$ & $m_{b,{\rm PS}}$(2~GeV) & $\Delta_{\rm th}$ & $\Delta_{\rm exp}$
& $\Delta_\alpha$ & $\Delta_{\rm tot}$& $\overline{m}_b$ \\
\hline
6 & 4460 & 40 & 50 & 35 & 70 & 4135 $\pm$ 65\\
8 & 4505 & 45 & 25 & 30 & 60 & 4170 $\pm$ 55\\
10 & 4515 & 45 & 15 & 25 & 55 & 4185 $\pm$ 50 \\
12 & 4520 & 45 & 10 & 20 & 50 & 4185 $\pm$ 45\\
14 & 4520 & 40  & 10 & 15  & 45 & 4185 $\pm$ 40\\
\hline
\hline
$n$ & $m_{b,{\rm RS}}$(2~GeV) & $\Delta_{\rm th}$ & $\Delta_{\rm exp}$
& $\Delta_\alpha$ & $\Delta_{\rm tot}$& $\overline{m}_b$\\
\hline
6 & 4315 & 55 & 50 & 25 & 80 & 4140 $\pm$ 70\\
8 & 4360 & 65 & 30 & 20 & 75 & 4180 $\pm$ 65\\
10 & 4370 & 65 & 20 & 10 & 70 & 4190 $\pm$ 60\\
12 & 4370 & 65 & 15 & 5 & 65 & 4190 $\pm$ 60\\
14 & 4370 & 65  & 10 & 5  & 65 & 4185 $\pm$ 55\\
\hline
\end{tabular}
\end{center}
 \caption{\it Extraction of $m_{b,{\rm PS/RS}}$(2~GeV) with errors for
 various $n$. All values are given in MeV and rounded to 5~MeV. The
 total error has been obtained by adding the partial errors in
 quadrature. The corresponding value for the $\MS$ mass with its error
 is given in the last column.\label{PStable}}
\end{table}

We summarize our results in Table~\ref{PStable}, where we also show
the combined error, $\Delta_{\rm tot}$, which is obtained by adding
the various errors in quadrature.  As expected, the experimental error
decreases with increasing $n$.  The results are all consistent with
each other, in particular if we take into account the additional
uncertainty mentioned above for $M_n$ with $n\le 8$, due to the
non-relativistic expansion in the energy integration.  Related to
this, we note that in computing the moments we do not use the exact
fixed-order coefficient at $\cO(\als^2)$, since we drop terms of
$\cO(\als^2/(\sqrt{n})^k)$ with $k\ge 1$. Again, this neglect is
potentially more of a problem for smaller moments. Let us also
reiterate that for too large values of $n$ the applicability of weak
coupling perturbation theory is questionable. We thus combine the
results of Table~\ref{PStable} by simply taking the value obtained by
the tenth moment
\begin{eqnarray}
\label{mPS2res}
m_{b,{\rm PS}}(2 {\rm GeV}) &=& 4.52 \pm 0.06 \;{\rm GeV}, \\
\label{mRS2res}
m_{b,{\rm RS}}(2 {\rm GeV}) &=& 4.37 \pm 0.07 \;{\rm GeV}.
\end{eqnarray}
Note that the PS value is consistent with the result of
Ref.~\cite{Beneke:1999fe}, but prefers smaller values for $m_b$ and
has a reduced error. 

Converting the PS and RS mass to the $\MS$-mass we obtain
$\overline{m}_b=4.19$~GeV with an error of 55~MeV and 60~MeV
respectively.  However, we also have to take into account the error in
the conversion itself. We consider two sources, the dependence of
$\overline{m}_b$ on the threshold mass used in the analysis and
second, the error due to missing higher-order corrections in the
conversion formula itself. To determine the first error, we start by
noting that the $\overline{m}_b$ values obtained with the PS and RS
scheme are very similar. We also extract the central value of
$m_{b,{\rm PS/RS}}$(1~GeV) for the moments and convert these results
to $\overline{m}_b$. These values of $\overline{m}_b$ differ at most
by around 20/15~MeV from the corresponding results obtained via
$m_{b,{\rm PS/RS}}$(2~GeV). To obtain an estimate for the error due to
missing higher-order corrections in the conversion formula we drop the
fourth order terms in the conversion and take as error the difference
in the value of $\overline{m}_b$ thus obtained. This error is about
10/5~MeV. We thus associate a total error of 20/15~MeV to the
conversion. If added in quadrature to the 55/60~MeV error, we obtain a
total error for $\overline{m}_b$ of around 60~MeV in both cases.

In conclusion, we have studied the effect of resumming logarithms for
non-relativistic sum rules. The logarithms turn out to be numerically
very important and improve the reliability of the theoretical
computation. This manifests itself in a reduced scale dependence and
an improvement of the convergence of the perturbative series. It
allows us to obtain an accurate value for the $\MS$ bottom quark mass
using a credible error estimate
\begin{equation}
\overline{m}_b(\overline{m}_b)=4.19\pm 0.06 \,{\rm GeV}.
\end{equation}
At this stage, the main problem appears to be the large size of the
perturbative corrections and to understand its origin. Further
improvements require the full NNLL computation of the sum rule,
especially the potentially large ultrasoft effects. Obviously, the
inclusion of all NNNLO effects will also be important and might lead
to a better control of the strong scale dependence for small values of
$\mu_s$ and the large size of the perturbative corrections.

\noindent
{\bf Acknowledgements}. AP acknowledges discussions with A.A. Penin
and J. Soto. AS thanks ECM Barcelona for hospitality during the course
of this work.


\end{document}